\documentclass[letterpaper, 10 pt, conference]{ieeeconf}  

\IEEEoverridecommandlockouts                              

\usepackage{amsmath,amssymb,amsfonts}
\usepackage{graphicx}
\usepackage{textcomp}
\usepackage{tabularx}
\usepackage{bytefield}
\usepackage{amsmath}
\usepackage{graphicx}
\usepackage{caption}
\usepackage{subcaption}
\usepackage{subcaption}
\usepackage{dblfloatfix} 
\graphicspath{{figures/}}
\usepackage{tkz-base}
\usepackage{tkz-euclide}
\usepackage[ruled,vlined]{algorithm2e}
\usepackage{comment}

\title{\LARGE \bf
Flowsim: A Modular Simulation Platform for Microscopic Behavior Analysis of City-Scale Connected Autonomous Vehicles
}

\author{Ye Tao$^{1}$, Ehsan Javanmardi$^{1}$, Jin Nakazato$^{1}$, Manabu Tsukada$^{1}$ and Hiroshi Esaki$^{1}$
    \thanks{*These research results were obtained from the commissioned research by the National Institute of Information and Communications Technology (NICT), JAPAN.}%
    \thanks{$^{1}$Graduate School of Information Science and Technology, The University of Tokyo, 1-1-1, Yayoi, Bunkyo-ku, Tokyo, 113-8657 Japan {\tt\small tydus@hongo.wide.ad.jp}}%
    \thanks{$^2$(The github URL will appear here)}%
}

\begin{document}

\maketitle
\thispagestyle{empty}
\pagestyle{empty}

\begin{abstract}

As connected autonomous vehicles (CAVs) become increasingly prevalent, there is a growing need for simulation platforms that can accurately evaluate CAV behavior in large-scale environments.
In this paper, we propose Flowsim, a novel simulator specifically designed to meet these requirements.
Flowsim offers a modular and extensible architecture that enables the analysis of CAV behaviors in large-scale scenarios.
It provides researchers with a customizable platform for studying CAV interactions, evaluating communication and networking protocols, assessing cybersecurity vulnerabilities, optimizing traffic management strategies, and developing and evaluating policies for CAV deployment.
Flowsim is implemented in pure Python in approximately 1,500 lines of code, making it highly readable, understandable, and easily modifiable.
We verified the functionality and performance of Flowsim via a series of experiments based on realistic traffic scenarios.
The results show the effectiveness of Flowsim in providing a flexible and powerful simulation environment for evaluating CAV behavior and data flow.
Flowsim is a valuable tool for researchers, policymakers, and industry professionals who are involved in the development, evaluation, and deployment of CAVs.
The code of Flowsim is publicly available on GitHub under the MIT license$^{2}$.

\end{abstract}

\section{INTRODUCTION}
\label{sec:introduction}

Road transportation has been one of the most essential services for human mobility since ancient times.
With recent technology advancements, connected autonomous vehicles (CAVs) have emerged as a promising solution to enhance road safety, efficiency, and convenience.
CAVs are vehicles equipped with advanced sensing, communication, and decision-making capabilities that enable them to operate autonomously and interact with their surroundings, including other vehicles, infrastructure, and pedestrians.

Comprehensive experiments and simulations are necessary to evaluate the performance and behavior of CAVs.
However, existing simulators often lack the necessary features and flexibility to accurately represent the behavior of CAVs' on-board units (OBUs).
Traffic simulators, such as SUMO\cite{Lopez2018-eq} focus on simulating traffic flow, whereas network simulators, such as NS-3\cite{Riley2010-te} and OMNET++\cite{Varga2010-vl} primarily model wireless radio communication and packet delivery.
VANET libraries and frameworks, such as artery\cite{Riebl2019-ou} and veins\cite{Sommer2011-vs}, provide the implementation of complete protocol stacks that are useful to further investigate the protocols within the network simulators.
Perception and localization simulators, such as Carla\cite{Dosovitskiy2017-so} and LGSVL\cite{Rong2020-ma}, provide realistic 3D environments, but they are not designed specifically for evaluating CAV behavior.
OpenCDA\cite{Xu2021-mn} is a cooperative driving automation simulator enabling a full procedure of perception, localization, planning, and actuating.
However, it only implements pre-defined cooperative driving models and is incapable of networking.
Kinematics simulators, such as MATLAB and Simulink, can evaluate the consequences of planning decisions.
However, they lack the capability to analyze complex CAV behaviors in a realistic context.
Thus, a new simulator that focuses on the behavior and data flow of CAVs' OBUs is urgently required.
Such a simulator should provide a city-scale experimental environment, offer high performance for large-scale scenarios, be extensible to accommodate various modules and functionalities, and provide flexibility for customizing and evaluating CAV behaviors.
Additionally, reproducibility is another crucial requirement for ensuring that experiments can be repeated by other researchers to verify the results.

In this study, we propose Flowsim, a novel simulator designed specifically to meet the requirements for evaluating CAV behavior and data flow.
Flowsim offers a modular and extensible architecture that enables the analysis of CAV behaviors in large-scale scenarios.
By providing a customizable simulation platform, Flowsim aims to accelerate the development and evaluation of cybersecurity measures and protocols for CAVs, thereby ensuring their safe and secure deployment.
In addition to its primary purpose of evaluating CAV behavior, Flowsim has a wide range of potential use cases, such as protocol evaluation, cybersecurity assessment, traffic optimization, and policy development.
The extensible and playable open environment of Flowsim fosters the exploration of novel algorithms and strategies for CAVs, thereby promoting innovation and collaboration within the research communities.

In this paper, we present a use case and quantitative evaluation of Flowsim, by assessing our previous work named Zero-Knowledge Proof of Traffic (zk-PoT).
Zk-PoT is a novel cryptographic scheme that enables vehicles to generate proofs of their observations to a specific target vehicle, while allowing for cross-verification of these proofs without compromising the privacy of any vehicle involved.
Our analysis of zk-PoT demonstrates its ability to prevent known attacks, enhancing overall security, efficiency, and data latency, but it needs to be supported by comprehensive quantitative evaluations.
Therefore, we constructed Flowsim, capable of simulating both the behaviors of proposed vehicles and potential attacks, allowing for comprehensive evaluation of zk-PoT's effectiveness and resilience.

The rest of this paper is organized as follows:
Section~\ref{sec:design} presents an overview of the design of Flowsim, highlighting its modular and extensible architecture.
Section~\ref{sec:implementation} discusses the implementation challenges and the solutions adopted in developing Flowsim.
In Section~\ref{sec:verification}, we verify the functionality and evaluate the performance of Flowsim through a series of experiments based on realistic traffic scenarios.
In Section~\ref{sec:ralated}, we discuss some topics of related works which could potentially be simulated using Flowsim.
Finally, Section~\ref{sec:conclusion} summarizes the contributions of Flowsim and outlines potential future directions for its development.

\section{DESIGN OF FLOWSIM}
\label{sec:design}
\begin{figure}[htb]
    \centering
    \includegraphics[width=\linewidth]{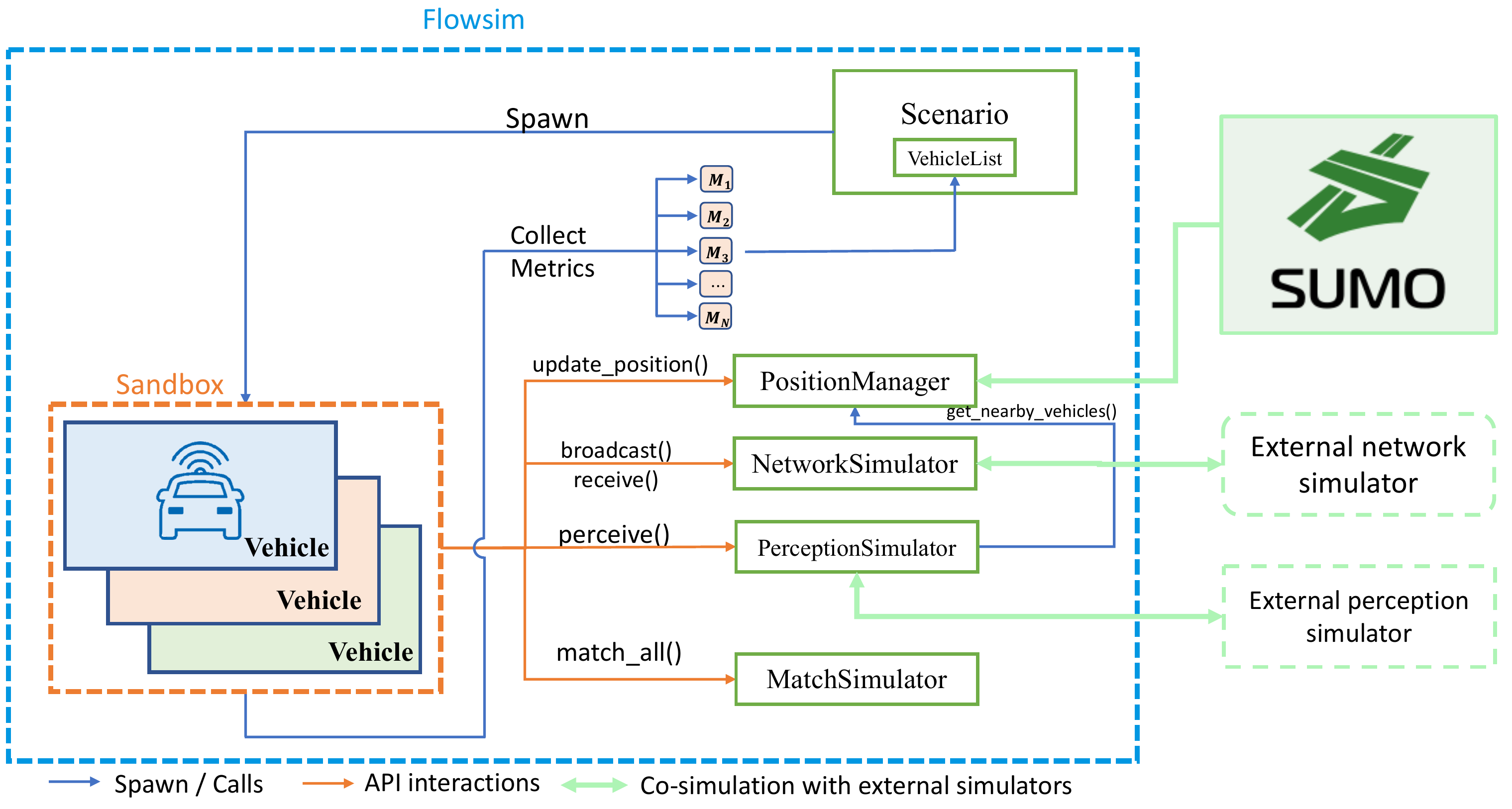}
    \caption{Architecture of Flowsim}
    \label{fig:flowsim}
\end{figure}

This section provides an overview of the structure and the design of Flowsim, which is a multi-agent simulator with a modular and extensible architecture.
Figure~\ref{fig:flowsim} illustrates the high-level structure of Flowsim, which can be divided into three main parts: simulators, sandbox, and vehicles within the sandbox.

\subsection{Scenario and Simulators}
As a simulation platform for evaluating the behavior of CAVs, Flowsim must provide agents with different types of information.
Therefore, Flowsim integrates various simulation mechanics, including positioning, perception, and networking.
\texttt{PositionManager} establish a co-simulation with SUMO and update the positions of all agents from SUMO.
Then, it provides the information to the relevant modules, such as the simulators and the vehicles.
Moreover, the position manager records the ground truths for external queries.
\texttt{NetworkSimulator} simulates networking-related functions.
Currently, it supports single-hop broadcast (SHB) in the ETSI GeoNetworking Standard\cite{Etsi2020-wk} as the only implemented networking feature.
\texttt{PerceptionSimulator} is a computer vision simulator.
Instead of rendering realistic scenarios and vehicles and performing object detection algorithms on video feeds, it extracts positional information from objects.
It simulates the projection from the ego vehicle's front camera, including FOV, occlusion, and filtering.
Finally, it outputs fully visible and recognizable numberplates based on predefined criteria.
Meanwhile, \texttt{MatchSimulator} is a specially designed ``non-standard'' module in Flowsim.
Generally, CAVs have two types of identity: numberplates like every vehicle, and an additional identity used for vehicle-to-everything (V2X) communication.
While matching these two identities is an open research topic in the literature\cite{Masuda2023-pa}, Flowsim simplifies the simulation setup by internally maintaining a match table using ground truth data from other simulators and the \texttt{VehicleList}.

The main purpose of Flowsim is to analyze and simulate the behavior and data flow within and among vehicles.
Currently, for simplicity and performance reasons, \texttt{NetworkSimulator}, \texttt{PerceptionSimulator}, and \texttt{MatchSimulator} are implemented from scratch.
However, if a more precise evaluation is required, such as simulating channel collisions in high-density traffic, the corresponding parts of Flowsim can be connected to external simulators.

Finally, the \texttt{Scenario} module handles the top-level information and controls the entire lifecycle of a simulation.
It spawns vehicles and simulator modules, manages ground truth data, and collects metrics from all simulators and agents.
By adopting a modular and extensible design, Flowsim provides a flexible and powerful simulation platform for analyzing CAV behavior and data flow.

\subsection{Sandbox and Different Types of Vehicles}

\begin{figure*}[hbtp]
    \centering
    \includegraphics[width=0.9\linewidth]{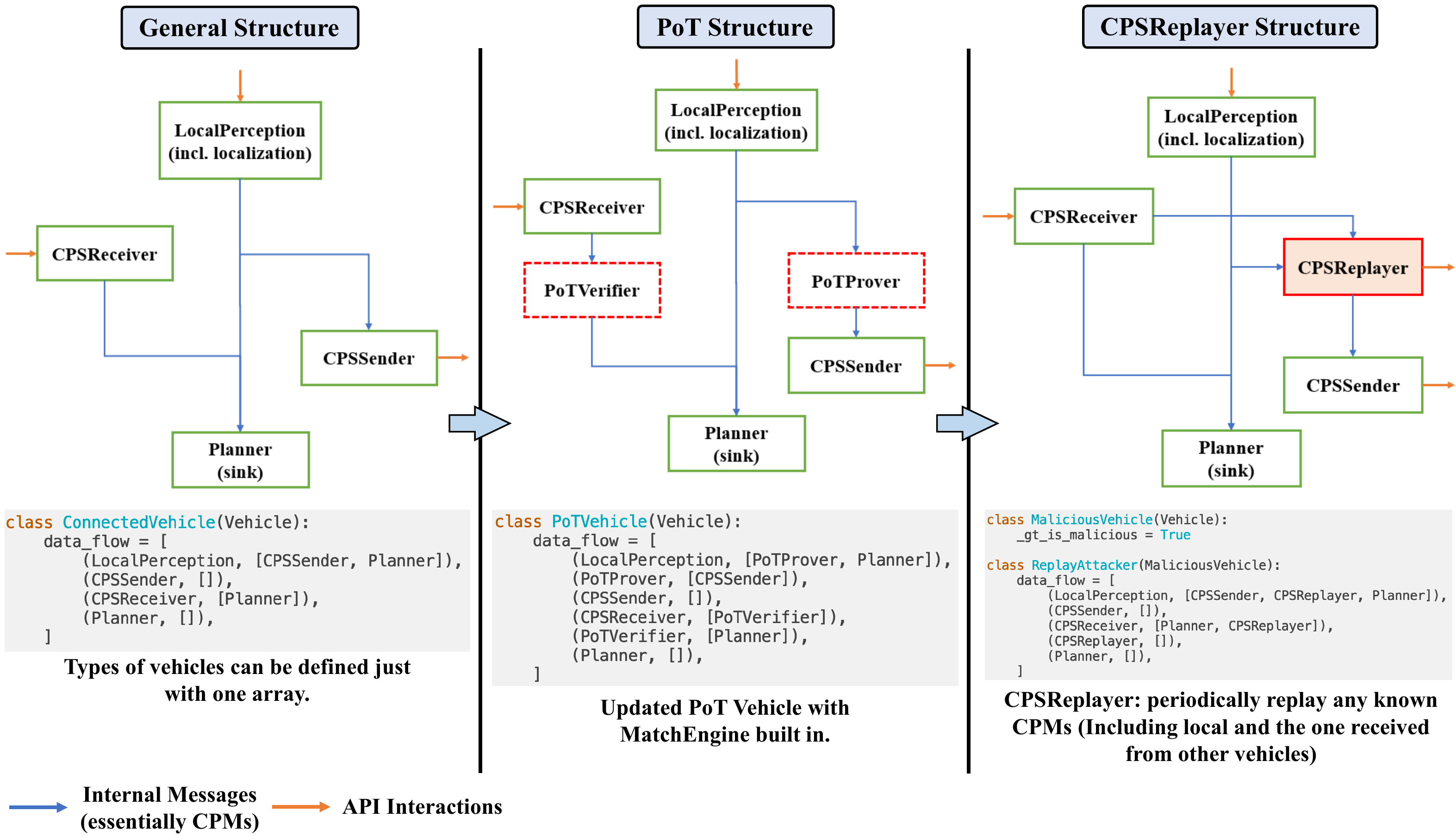}
    \caption{Example of defining vehicle types}
    \label{fig:flowsim-defining-vehicle-types}
\end{figure*}

\begin{table*}[bt]
    \centering
    \caption{Implemented vehicle types in Flowsim}
    \label{tab:flowsim-vehicle-types}
    \begin{tabular}{rl}
        \hline
        Name & Description \\
        \hline
        UnconnectedVehicle & Traditional vehicles with perception, but not connected to V2X \\
        ConnectedVehicle   & Perception and V2X communication capable vehicles, can send and receive CPMs \\
        PoTVehicle         & Proof of Traffic capable vehicles, can send and verify proofs \\
        SpamAttacker       & Attacker vehicles sending random fake objects to make the road look congested \\
        ReplayAttacker     & Attackers replaying CPM of both received and local perceived objects to confuse or overload other vehicles \\
        SilenceAttacker    & Selfish vehicles that only listen to V2V communications without contributing anything \\
        \hline
    \end{tabular}
\end{table*}

To ensure the legitimacy of vehicle implementations, Flowsim incorporates a sandbox environment that explicitly controls the information accessible to agents and the operations they can perform.
Various types of vehicles have been implemented in the sandbox.
While the current focus is on vehicles, Flowsim can easily accommodate other types of agents with V2X capabilities, such as roadside units (RSUs), bicycles, and pedestrians with smartphones.

Currently, Flowsim supports several types of vehicles, as listed in Table \ref{tab:flowsim-vehicle-types}.
These include general-purpose vehicles like traditional unconnected autonomous vehicles, V2X connected vehicles, and the Proof-of-Traffic (PoT) vehicles\cite{Tao2023-nn}.
Since cybersecurity is a key concern in the PoT context, Flowsim also implements different types of attacker vehicles with ``tampered'' On-Board Unit (OBU) implementations, including spammers, replay attackers, and selfish attackers.

Flowsim also allows for the easy implementation of other vehicle types.
For example, a human-driven vehicle can be represented as a dummy vehicle, which is equivalent to an empty OBU.
By providing this range of vehicle types within a sandbox environment, Flowsim offers flexibility for simulating various scenarios and studying different aspects of cooperative autonomous systems.

\subsection{Flow Inside the Vehicles}
The design inside each vehicle follows a flow-based approach, which inspired the name \textbf{Flowsim}.
The vehicle functions are abstracted into atomic and autonomous modules.
Each module takes its input data, processes them, and produces the output data.
Additionally, the modules can interact with the simulators outside of the sandbox by calling the explicitly provided APIs.

The data flow among the vehicle modules is represented by a Directed Acyclic Graph (DAG).
In contrast to the widely-used Pub-Sub design pattern, we explicitly define and implement the data flow, using the adjacency list representation of the DAG.
This explicit representation offers several advantages, including higher performance, easier design and debugging, and the elimination of the risk of infinite loops.

Furthermore, the data passing through the DAG are isomorphic, which means that all data transfers within a vehicle essentially consist of Collective Perception Messages (CPMs)\cite{Etsi2023-ac} with private extensions.
If required, some modules can call NetworkSimulator to broadcast those CPMs onto the network, after cleaning up the internally used private extensions.

Leveraging the strength and flexibility of Python, we developed a Domain Specific Language (DSL) that represents the adjacency list and enables the definition of different vehicle types using this DSL.
As depicted in Figure \ref{fig:flowsim-defining-vehicle-types}, defining a vehicle based on a specific data flow is as simple as listing the module names and specifying the next modules to which the output of each module should flow.
This flexibility enables us to manipulate vehicle types easily, including creating various types of attackers.
By adopting a flow-based design approach, Flowsim provides a powerful and customizable framework for modeling and analyzing the behavior and interactions of vehicles within a simulation environment.
\subsection{Metric Collection}
Depending on the requirements, Flowsim enables the collection of various metrics after each tick of the simulation.
The \texttt{MetricCollector} module serves as the utility for metric collection.
Researchers can customize their implementation to produce metrics in formats that match their requirements, such as text, CSV, JSON, or DataFrame.
In the current implementation, the \texttt{MetricCollector} serializes the metrics of each tick into separate lines of JSON.
Moreover, it generates an ``index'' file that records the position of each line in the original file for rapid seeking.

Currently, Flowsim records four sets of metrics, three of which are generally applicable.

\subsubsection{\textbf{bytes\_sent}} tracks the number of bytes sent by each vehicle in each tick, allowing for the calculation of bandwidth consumption.

\subsubsection{\textbf{local/received/all\_objects}} provide insights into the accumulated number of objects collected by each vehicle from different sources.
These include the objects perceived by sensors, objects received from cooperative perception, and the total number of objects available to the planner.
These statistics are crucial for evaluating the performances and efficiencies of CAVs.

\subsubsection{\textbf{time\_to\_verify\_histogram}} measures the distribution of the delay from receiving a new object to ``believing'' the data associated with it.
This is specifically relevant to security-related research and can be used to compare the timeliness of different methods.

By collecting these metrics, Flowsim enables researchers to gain valuable insights into various aspects of the simulation, ranging from network performance to the behavior and efficiency of CAVs, as well as the timeliness of security-related processes.

\section{IMPLEMENTATION DETAILS}
\label{sec:implementation}

In this section, we will discuss the crucial implementation details that can impact the performance of Flowsim and our corresponding solutions.

\subsection{Programming language and its implementation}

Python was selected as the programming language for Flowsim because of widespread use among researchers and specialties for rapid prototyping.
To maintain maximum readability, Flowsim was implemented in pure Python.
Given the flexibility requirement and modular design, it was infeasible to write most parts in C extensions.
Therefore, we selected the PyPy implementation.

PyPy is a Just-In-Time (JIT)-enabled Python implementation that offers several times faster execution than the official CPython implementation in most cases.
It achieves all this under a hood, maintains full compatibility with the official CPython, and does not require modifications to the Python code.
However, PyPy has some limitations, particularly when invoking C extensions.
Calling C code interrupts the JIT compilation and significantly hampers performance, making it even worse than the CPython interpreter.
This limitation is the main drawback of PyPy; therefore, it is crucial to minimize the time spent calling C code, which includes libraries such as \texttt{numpy} and interactions with \texttt{SUMO}.

It is important to note that despite its advantages, PyPy is not a solution to overcome the limitations imposed by the Global Interpreter Lock (GIL).
The GIL still affects the performance of multi-threading in PyPy, rendering it inefficient for true parallel execution.
To address the limitations imposed by the GIL, we have considered the possibility of employing the multiprocessing approach in Flowsim.
It involves spawning multiple processes and facilitating communication among them to realize parallel execution.
Although we have not yet implemented multiprocessing in Flowsim, it is a potential solution for enhancing performance.

Switching to multiprocessing typically requires a significant program redesign.
However, the sandbox design of Flowsim isolates the vehicles and defines explicit APIs between vehicles and simulators; thereby minimizing the need for extensive modifications.
If we were to implement multiprocessing, we would spawn worker processes for vehicle agents and utilize Remote Procedure Calls (RPCs) to wrap API calls to the simulators.
This approach enables the efficient utilization of system resources and can boost the overall performance of Flowsim.

By leveraging the PyPy implementation and considering the potential use of multiprocessing, Flowsim strikes a balance between performance optimization and code maintainability, enabling researchers to perform simulations efficiently while enjoying the readability of Python.

\subsection{PositionManagerV2}
PositionManagerV2 is the current implementation of the Position Manager in Flowsim.
It is responsible for synchronizing the list of vehicles and their position information with the external traffic provider SUMO.
One of the key functions of the Position Manager is \verb|get_nearby_vehicles()|.
The function checks all the vehicles on the map and returns those that are within a predefined distance from the egovehicle.
In the na\"ive implementation, iterating over all vehicles results in a time complexity of $O(N)$, where $N$ indicates the total number of vehicles.

To optimize this process, we introduce an algorithm that divides the map into orthogonal grids of a predefined size.
Each vehicle is then assigned to a specific grid cell according to its current position.
This approach allowed us to answer each query efficiently by filtering through the egovehicle's cell and its eight adjacent cells.
Using this optimization, we achieved a time complexity of $O(n)$ per query, where $n$ represents the number of vehicles in each cell.
The preprocessing step requires $O(N)$ time to assign vehicles to the grid cells.

To handle corner cases, such as when the coordinate becomes -1.6 because the centerline of the edge road is 0, we surround the map grids with margins of two cells on each side.
Although using hexagonal grids could offer slightly better efficiency, we opted for orthogonal grids for the sake of simplicity and better readability of the implementation.

By employing this optimized algorithm for querying nearby vehicles, PositionManagerV2 significantly improves the performance of Flowsim, enabling efficient and accurate retrieval of vehicles within a specified distance from the egovehicle.

\subsection{PerceptionSimulatorV3}
PerceptionSimulatorV3 is the current implementation of the perception simulator in Flowsim.
In this version, we utilized a simple computer vision model to simulate perception capabilities.
The key features of this model are as follows:

\subsubsection{\textbf{Two-dimensional world}} The perception simulator assumes a world with only two dimensions, disregarding the altitude or Z-axis.

\subsubsection{\textbf{Field of view (FOV)}} The field of view of the camera is thus reduced to a one-dimensional representation, presented as an angle range.

\subsubsection{\textbf{Simplified vehicle shapes}} All vehicles in the simulation are represented as perfect rectangles to facilitate easier geometric calculations.

\subsubsection{\textbf{Numberplate placement}} Numberplates are simulated as shorter lines located at the center of the front and rear bumpers of the vehicles, thus they are centrosymmetric about the center of the vehicle.

\subsubsection{\textbf{Exclusion of polygons}} This implementation does not consider polygons such as buildings in the simulation.

\setcounter{subsubsection}{0}

Although we have simplified the perception algorithm to be purely geometric-based, it remains the most computationally intensive task within the simulator.
The perception algorithm in PerceptionSimulatorV3 can be implemented in the following five steps:

\subsubsection{\textbf{Vehicle Reconstruction}}

In Flowsim, we reconstructed the dimensions of the vehicles based on the position information provided by SUMO.
We accurately determine the dimensions and shape of the vehicles in the simulation by utilizing the position of the center of the front bumper and the heading of each vehicle.
The reconstruction process ensures realistic representations of the vehicles and enables accurate interactions and behaviors within the simulated environment.

\subsubsection{\textbf{Coordinate conversion and FOV filtering}}

The next step involves converting the retrieved vehicle boxes from the original ``world'' coordinate system into ``camera'', a Cartesian coordinate system originating from the front camera of the egovehicle.
Specifically, the camera of the egovehicle is located at $(0,0)$ and is directed towards the positive x-axis.

Let us consider the original ``world'' coordinate system with the egovehicle's camera located at $(x_0, y_0)$ with a heading of $\beta_0$.
If a vehicle is located at $(x, y)$ with a heading of $\beta$ in this coordinate system, it can be projected onto the new coordinate system $[x', y', \beta']$ using Equation~\ref{equ:translation_matrix}.

\begin{equation}
    \begin{bmatrix}
        x' \\
        y' \\
        \beta'
    \end{bmatrix}
    =
    \begin{bmatrix}
        \cos(\beta_0) & \sin(\beta_0) & 0\\
        -\sin(\beta_0) & \cos(\beta_0) & 0\\
        0 & 0 & 1
    \end{bmatrix}
    \begin{bmatrix}
        x - x_0 \\
        y - y_0 \\
        \beta - \beta_0
    \end{bmatrix}
    \label{equ:translation_matrix}
\end{equation}

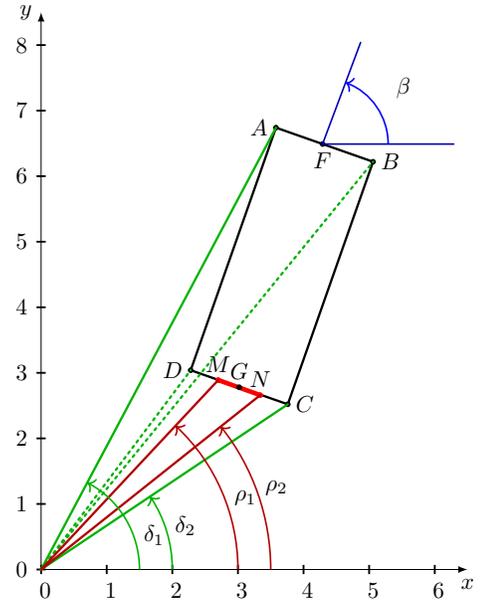
\begin{figure}[htbp]
    \centering
    \resizebox{.75\linewidth}{!}{
        \begin{tikzpicture}
            \tkzInit[xmax=6,xmin=0,ymax=8,ymin=0]
            \tkzAxeXY
            \tkzDefPoint[label=left :$A$](3.58,6.74){A}
            \tkzDefPoint[label=right:$B$](5.06,6.22){B}
            \tkzDefPoint[label=right:$C$](3.76,2.52){C}
            \tkzDefPoint[label=left :$D$](2.28,3.04){D}
            \tkzDefPoint[label=below:$F$](4.29,6.49){F}
            \tkzDefPoint(4.29,7.49){Fy}
            \tkzDefPoint(4.29,6.49){Fb}
            \tkzDefPoint[label=above:$G$](3.02,2.78){G}
            \tkzDefPoint[label=above:$M$](2.70,2.89){M}
            \tkzDefPoint[label=above:$N$](3.34,2.66){N}
            \tkzDefPoint(0,0){O}
            \tkzDefPoint(1,0){x}
            
            \tkzDrawSegments[color=black!70!black,line width=1pt](A,B B,C C,D D,A)
            \tkzDrawSegments[color=red!70!red,line width=2pt](M,N)
            \tkzDrawPoints[color=black!70!black](A,B,C,D,F,G)
            \tkzDrawPoints[color=red!70!red](M,N)
            \tkzDrawSegments[color=green!70!black,line width=1pt](O,A O,C) 
            \tkzDrawSegments[color=green!70!black,line width=1pt,dotted](O,B O,D) 
            \tkzDrawSegments[color=red!70!black,line width=1pt](O,M O,N) 
    
            \tkzDefPoint(6.29,6.49){Fx}
            \tkzDefPoint(4.87,8.04){Fb}
            \tkzDrawSegments[color=blue!70!black,line width=0.7pt](F,Fx F,Fb)
            \tkzMarkAngle[color=blue!70!blue,line width=0.7pt,arrows=->](Fx,F,Fb)
            \tkzLabelAngle[pos=1.5](Fx,F,Fb){$\beta$}
    
            \tkzMarkAngle[color=green!70!black,size=1.5,line width=0.7pt,arrows=->](x,O,A)
            \tkzLabelAngle[pos=1.8](x,O,C){$\delta_1$} 
            \tkzMarkAngle[color=green!70!black,size=2.0,line width=0.7pt,arrows=->](x,O,C)
            \tkzLabelAngle[pos=2.3](x,O,C){$\delta_2$}
    
            \tkzMarkAngle[color=red!70!black,size=3.0,line width=0.7pt,arrows=->](x,O,M)
            \tkzLabelAngle[pos=3.3](x,O,N){$\rho_1$}
            \tkzMarkAngle[color=red!70!black,size=3.5,line width=0.7pt,arrows=->](x,O,N)
            \tkzLabelAngle[pos=3.8](x,O,N){$\rho_2$}
    
        \end{tikzpicture}
    }
    \caption{
    Relevant points and corresponding angles in the camera coordinate. \\
    Green lines: projections of corners of vehicles; \\
    Red lines: projections of the rear numberplate; \\
    $\beta$: heading angle related to the camera position; \\
    $\delta_1$, $\delta_2$: argument angles of corners of vehicles; \\
    $\rho_1$, $\rho_2$: argument angles of the rear numberplate; \\
    }
    \label{fig:relevant-points}
\end{figure}

Once the vehicles are converted into a new coordinate system, it becomes simple and straightforward to filter out the irrelevant vehicles based on the camera's FOV.
In Figure~\ref{fig:relevant-points}, we can observe that a vehicle is considered irrelevant only if NONE of its corners, denoted as $A, B, C, D$, fall within the FOV.
In other words, these vehicles have no chance of being identified or of occluding other vehicles.

\subsubsection{\textbf{Calculate the Projection}}

The centrosymmetric assumption of the vehicles and their numberplates allowed us to normalize the ``heading'' direction of the vehicle with respect to the camera.
Specifically, if $|\beta|$ (the absolute value of the heading) is greater than $\pi / 2$, the vehicles are rotated by $\pi$ radians.
This normalization ensures that the vehicles are always effectively ``tailing'' the camera, allowing us to focus solely on their (equivalent) rear numberplates.

Referring to Figure~\ref{fig:relevant-points}, we can narrow our focus to six key points: corners $A, B, C, D$, and the ends of the numberplates, denoted as $M$ and $N$.

Since the FOV of the camera is represented by angles, we can simplify the calculations by transforming the coordinates once again into polar coordinates.
However, for efficiency purposes, we do not need a complete polar coordinate representation of the points.
Instead, we only require the arguments (i.e., the angle from the positive x-axis to the point).

Thus, we can calculate the range of the vehicle box ($\delta_1, \delta_2$) and the range of its numberplate ($\rho_1, \rho_2$) using Equation~\ref{equ:delta_rho}.

\begin{equation}
\begin{aligned}
\delta_1 &= \min(Arg(A), Arg(B), Arg(C), Arg(D)) \\
\delta_2 &= \max(Arg(A), Arg(B), Arg(C), Arg(D)) \\
\rho_1   &= \min(Arg(M), Arg(N)) \\
\rho_2   &= \max(Arg(M), Arg(N))
\end{aligned}
\label{equ:delta_rho}
\end{equation}

$Arg(P)$ gives the argument of a point $P$ ranging from $-\pi$ to $\pi$, where $min$ and $max$ denote getting the minimum and maximum value, respectively.

Subsequently, we sorted the candidate vehicles based on their distance from the center of the rear bumper, denoted as $G$.

\subsubsection{\textbf{Occlusion Calculation}}

Given that the current implementation ignores the polygons from the map, the only factor to consider is the occlusion between two vehicles.

To facilitate occlusion calculation, we mapped the argument angles $\delta_1$, $\delta_2$, $\rho_1$, $\rho_2$ within the polar coordinate system onto a ``flattened'' number line ranging from $-\pi$ to $\pi$.
This mapping allows us to transform the problem of ``occlusion'' into an interval coverage problem.

\begin{figure*}[b]
    \begin{subfigure}[b]{0.3\linewidth}
        \centering
        \includegraphics[width=\linewidth]{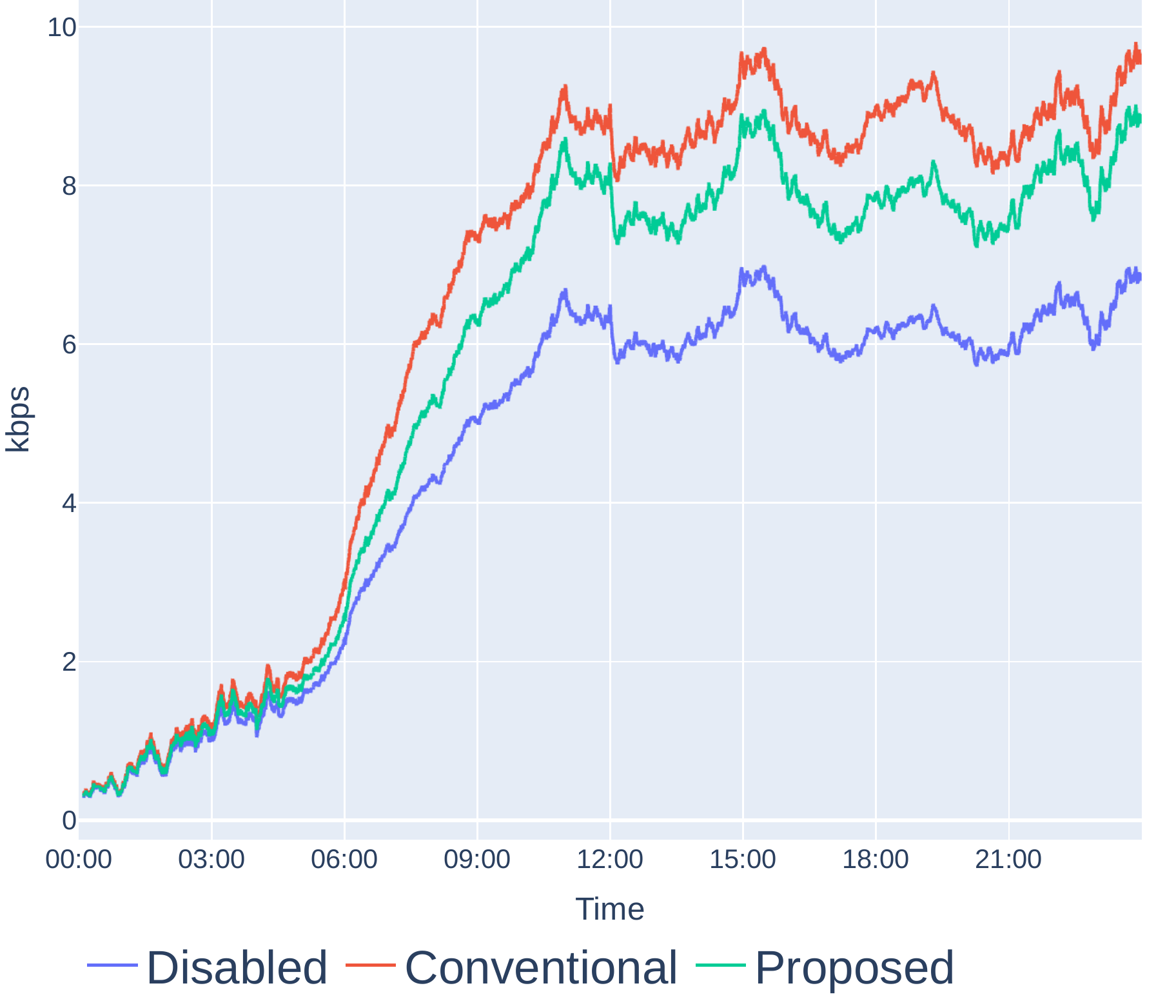}
        \caption{Line: average bandwidth consumption}
        \label{fig:example-figures-line}
    \end{subfigure}
    \hfill
    \begin{subfigure}[b]{0.3\linewidth}
        \centering
        \includegraphics[width=\linewidth]{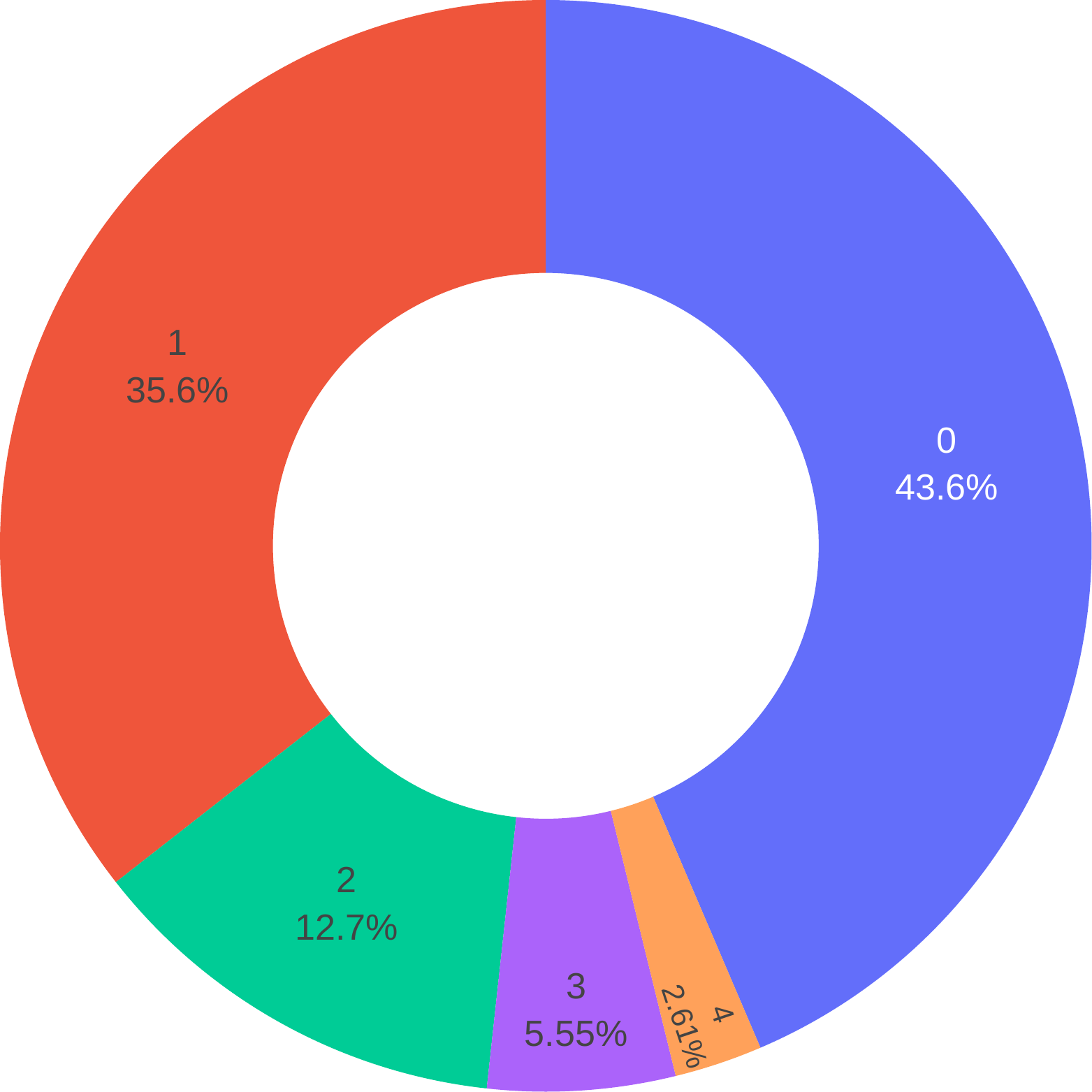}
        \caption{Pie: Time-to-Verify (TTV) distribution}
        \label{fig:example-figures-pie}
    \end{subfigure}
    \hfill
    \begin{subfigure}[b]{0.35\linewidth}
        \centering
        \includegraphics[width=\linewidth]{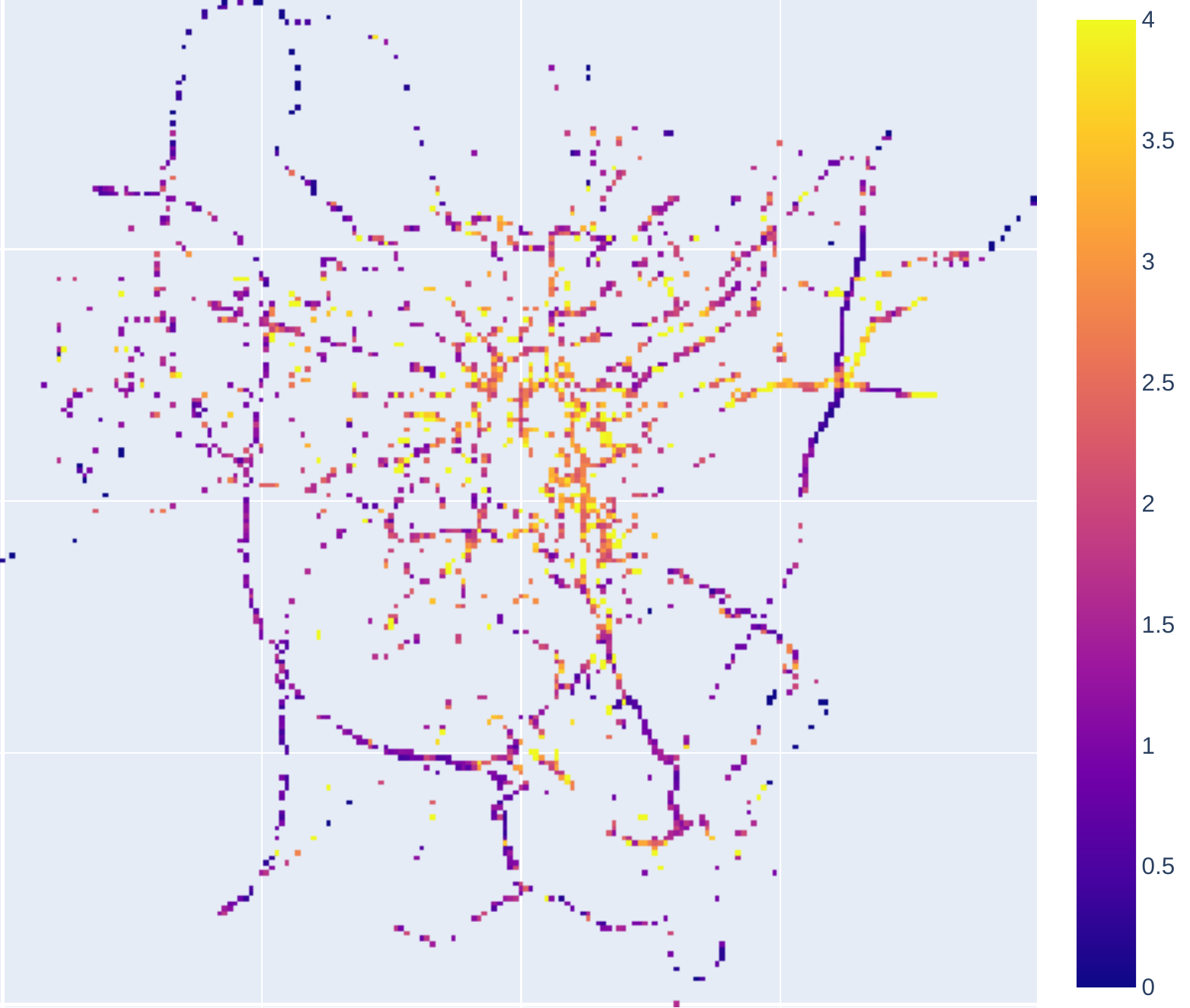}
        \caption{Heatmap: cooperative perception ratio}
        \label{fig:example-figures-heatmap}
    \end{subfigure}
    
    \caption{Example figure types generated from collected metrics}
    \label{fig:example-figures}
\end{figure*}

The algorithm for this step is defined as Algorithm~\ref{alg:get_visible_lines}.
By leveraging the segment tree\cite{De_Berg2008-yt} data structure and employing the discretization method, the time complexity can be reduced from $O(|C|^2)$ to $O(|C| \log |C|)$, where $|C|$ represents the number of candidate vehicles.

\begin{algorithm}[h]
  \SetAlgoLined
  \SetKwInOut{Input}{Input}
  \SetKwInOut{Output}{Output}

  \Input{$C$ as candidate vehicles in projection format, sorted by the distance of $G$}
  \Output{Visible vehicles in projection format}

  \BlankLine

  \If{$C = \varnothing$}{
    \Return $\varnothing$\;
  }

  $S \gets \varnothing$\;
  $R \gets [C^0]$\;
  
  \For{$(\ell, m) \text{ in } \text{zip}(C[:-1], C[1:])$}{
    $S \gets S \cup \{\delta_1^\ell, \delta_2^\ell\}$\;
    \If{$(\rho_1^m, \rho_2^m) \notin S$}{
      $R \gets R \cup [m]$\;
    }
  }

  \Return $R$\;

  \caption{get\_visible\_lines}
  \label{alg:get_visible_lines}
\end{algorithm}

\subsubsection{\textbf{Filter by Heading}}
Finally, even if the numberplate of a vehicle is not occluded, it may still be considered unrecognizable if it is rotated too much.
Therefore, we filtered the candidate vehicles again according to their headings, which is $\beta$ in the camera coordinate system.
This filter ensures that only vehicles within an acceptable rotation range are considered identifiable.

By going through these steps, we can ensure that the remaining vehicles can be perceived and identified by the camera of the egovehicle.
The PerceptionSimulatorV3 records the ground truth and returns the corresponding numberplates to the egovehicle.

\section{Verification}
\label{sec:verification}
In this section, we verify the functionality and performance of Flowsim based on specific settings.
It is important to note that Flowsim was initially developed to evaluate the work Zero-Knowledge Proof of Traffic (zk-PoT)\cite{Tao2023-nn}.
As a result, certain implementations and experiment parameters were tailored for this scenario.
For the experiments, we selected a 24-hour realistic traffic dataset generated from the road network of Luxembourg, known as the LuST scenario\cite{Codeca2015-wh}.
The LuST scenario includes 280,000 distinct vehicles that have been part of the scenario at some point.
During the peak hour at 19:00, there are over 8,000 active vehicle agents to be simulated.
To maintain a satisfactory level of simulation accuracy, the tick resolution was set to 1 second per tick.
Consequently, the full-sized LuST scenario can be completed in less than one day of wall time.

Throughout the experiments, we collected various metrics per tick per vehicle, including the positions of the vehicles, the number of bytes sent, and the number of known objects from different modules.
This comprehensive data collection approach provides maximum flexibility for data analysis, enabling us to examine the required information and generate relevant visualizations.
Figure~\ref{fig:example-figures} shows examples of the data aggregated from the collected metrics.

Figure~\ref{fig:example-figures-line} shows a line chart depicting the average bandwidth consumption over the simulation time.
The data enables us to assess the impact of the proposed method on the network.
It was obtained by calculating the mean value of all the vehicles for each tick.

Figure~\ref{fig:example-figures-pie} presents a pie chart representing the Time-to-Verify (TTV) for a given tick.
TTV indicates the number of seconds of delay before the newly received vehicle data are protected by cross-verification.
This metric measures the timeliness of the proposed method and can be compared with conventional statistic-based approaches.
It was obtained by summing up the TTV histogram data for each vehicle activated by the tick.

Figure~\ref{fig:example-figures-heatmap} shows a heatmap of Cooperative Perception Ratio (CPR) of our work at 19:00.
CPR reflects the contribution level of information received from other vehicles compared with information gathered from local perceptions.
The metric is calculated by dividing the sum of the objects not perceived locally by the sum of the locally perceived objects.

The experiments were conducted on a desktop PC equipped with an Intel i9-13900kf CPU and 128GB DRAM.

\begin{figure}[htbp]
    \centering
    \includegraphics[width=\linewidth]{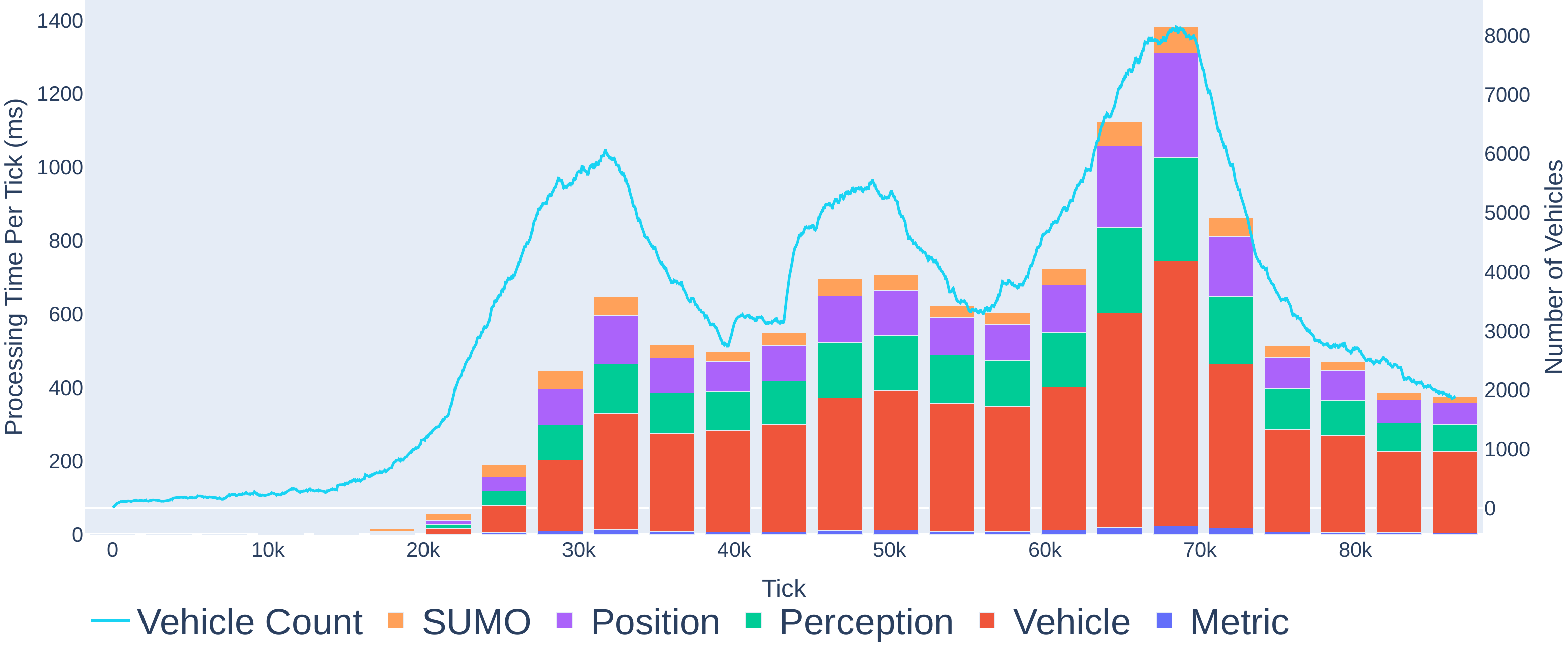}
    \caption{Number of vehicles and processing time breakdown}
    \label{fig:time-breakdown}
\end{figure}

We timed the common scenario, in which all vehicles were capable of cooperative perception.
Figure~\ref{fig:time-breakdown} illustrates the total processing time per tick and provides a breakdown of the simulators and agents.
A maximum processing time of 1.4 seconds per tick was observed when the number of active vehicles reached 8,200.
Most ticks were processed in approximately 600 milliseconds, which indicates an average processing time of approximately 15 microseconds per vehicle.

\section{RELATED WORK}
\label{sec:ralated}
Thanks to its extensible and flexible nature, Flowsim exhibits the capability to be modified and extended, thereby accommodating a diverse range of use cases. These use cases encompass various aspects, as outlined below:

\subsubsection{\textbf{Behavioral Analysis}} Flowsim enables the study and analysis of the behavior of CAVs, providing insights into their interactions, driving policies, and overall behavior in different traffic conditions.
\subsubsection{\textbf{Protocol Evaluation}} Flowsim facilitates the evaluation of communication and networking protocols for CAVs, assessing their performance, safety, and reliability in CAV networks.
\subsubsection{\textbf{Cybersecurity Assessment}} Flowsim supports the assessment of cybersecurity vulnerabilities in CAVs, enabling researchers to simulate attack scenarios, evaluate security measures, and develop robust countermeasures.
\subsubsection{\textbf{Traffic Optimization}} Flowsim provides a platform for evaluating and optimizing traffic management strategies, allowing researchers to test traffic control algorithms, congestion management, and route planning techniques for improving traffic flow and reducing travel times.
\subsubsection{\textbf{Policy Development}} Flowsim aids in the development and evaluation of policies and regulations for CAVs, enabling simulation-based assessments of various interventions, such as lane assignment, speed limits, and priority rules on traffic flow, safety, and energy efficiency.

Among others, we delve into two specific use cases in detail: misbehavior detection and pseudonym changing strategies.
These examples highlight the practical application of Flowsim in addressing key challenges in the domain of CAVs and provide a deeper understanding of its capabilities and effectiveness.

\subsection{Misbehavior detection}
Misbehavior Detection (MBD) is a key research area for CAVs, which focuses on monitoring the semantics of V2X messages to identify potential misbehaving entities.

In~\cite{Tsukada2022-ze}, the authors propose a method to enhance MBD by leveraging observation data from other vehicles via CPMs.
The researchers conducted experiments using realistic traffic scenarios and achieved promising results: it reduced the rejection rate of valid CAMs (false positives) by 15\% in the scenario with high traffic density while maintaining the rate of correctly detecting attacks (true positives).
The work was originally evaluated in a small-scale scenario using Carla.
It is easy to rewrite the proposed method using Flowsim, to achieve a simulation in a city-scale environment to evaluate the impacts, for example, of remote vehicles.

\cite{Kamel2020-gz} introduces a simulation framework called Framework for Misbehavior Detection (F2MD), as a tool for the research community to develop, test, and compare MBD algorithms.
The framework includes a comprehensive list of attacks, basic and advanced detection algorithms, a Python/C++ bridge for importing artificial intelligence algorithms, basic Pseudonym Change Policies, and a visualization tool for real-time performance analysis of the MBD system.
F2MD primarily focuses on networking, which made it highly suitable to be co-simulated with Flowsim, which takes care of other factors like vehicle position, identification, and perception.

\subsection{Pseudonym changing strategies}
Pseudonyms and their changing strategies are crucial for CAVs operating in Vehicular Ad-hoc Networks (VANETs) to protect the privacy and anonymity of vehicles, preventing tracking and unauthorized access to sensitive information such as location data, thus ensuring the security and trustworthiness of the overall system.

MixGroup\cite{Yu2016-ml} is a new privacy-preserving scheme that leverages the sparse meeting opportunities to efficiently exchange pseudonyms.
By integrating the group signature mechanism, MixGroup creates extended pseudonym-changing regions where vehicles can successively exchange their pseudonyms.
The performance is evaluated by NS-3 and SUMO, which shows favorable performance even in low traffic conditions.

Estimation of Neighbors Position privacy scheme with an Adaptive Beaconing approach (ENeP-AB)\cite{Zidani2018-fn} is a location privacy scheme that preserves the Quality of Service (QoS) of road safety applications in VANETs, by changing pseudonyms based on the number of neighbors and their predicted positions.
In the simulation, they used PREXT\cite{Emara2016-hr}, an extension to the Veins framework, which itself is based on OMNET++\cite{Varga2010-vl}.
The results show that the beaconing rate approaches create confusion in location and time during pseudonym changes, thus improving the safety.

Dynamic Mix-zone for Location Privacy (DMLP)\cite{Ying2013-ev} is another work preserving location privacy that features dynamic on-demand creation of mix-zones.
The authors only conduct simulations on a suburban road with two lanes using a vehicle generation that follows Poisson distribution.
The protocol can be implemented based on Flowsim and evaluated on a larger scale and realistic traffic patterns.


\section{CONCLUSION}
\label{sec:conclusion}
In this study, we presented Flowsim, a novel simulator designed to evaluate the behavior and data flow of CAVs.
Recognizing the limitations of existing simulators in capturing the intricacies of CAV behaviors, Flowsim was developed with a focus on modularity, extensibility, and performance to address the unique requirements of behavior evaluation.
We introduced the motivation behind Flowsim, highlighting the need for a simulator that specifically targets the behavior of CAVs.
Existing simulators are useful for various aspects of transportation and network simulation, but they often fail to provide a comprehensive evaluation platform for CAV behaviors.
Flowsim fills this gap by offering a city-scale experimental environment, high-performance simulation capabilities, and a modular architecture that can be customized and extended to suit specific research needs.

Through carefully designed experiments, we assessed the functionality and performance of Flowsim.
These experiments ensure the ability of Flowsim to capture realistic CAV behaviors, evaluate the effectiveness of cybersecurity measures, and assess the impact factors on CAV performance.

Flowsim is written in merely 1500 lines of pure Python code, making it highly readable, understandable, and easily modifiable for different applications.
We not only welcome researchers to use Flowsim for their studies but also strive to create a learnable and playable open environment.
By providing accessible and well-documented code, we aim to foster collaboration and knowledge sharing within the research community.
The code of Flowsim is available in the largest opensource community GitHub.
Any contributions and pull requests are highly welcome.

In future work, we aim to further enhance Flowsim by integrating more advanced simulation components, such as advanced perception models and detailed communication protocols, and implementing planning and control algorithms.
Moreover, we plan to explore the application of Flowsim in various research fields, including traffic optimization, intelligent transportation systems, and cooperative driving scenarios.
By continuously improving and expanding Flowsim's capabilities, we endeavor to contribute to the advancement of CAVs and foster safer and more efficient transportation systems.

\section*{ACKNOWLEDGEMENT}
These research results were obtained from the commissioned research Grant number 01101 by the National Institute of Information and Communications Technology (NICT), Japan. This work was partly supported by JSPS KAKENHI (grant numbers: 21H03423).

\bibliographystyle{IEEEtran}
\bibliography{IEEEabrv,root}

\end{document}